# FORS-Up: Making the most versatile instrument in Paranal ready for 15 more years of operations

H. M. J. Boffin[*a], F. Derie[a], A. Manescau[a], R. Siebenmorgen[a], V. Baldini[b], G. Calderone[b], R. Cirami[b], I. Coretti[b], P. Di Marcantonio[b], J. Kolsmanski[a], P. Lilley[a], S. Moehler[a], M. Nonino[b], G. Rupprecht[a], A. Silber[a]
[a] ESO, Karl-Schwarzschild-str. 2, 85748 Garching bei Munchen, Germany
[b] INAF-OATs, Trieste, Italy

## ABSTRACT

The FORS Upgrade project (FORS-Up) aims at bringing a new life to the highly demanded workhorse instrument attached to ESO's Very Large Telescope (VLT). FORS2 is a multimode optical instrument, which started regular science operations in 2000 and since then, together with its twin, FORS1, has been one of the most demanded and most productive instruments of the VLT. In order to ensure that a FORS shall remain operational for at least another 15 years, an upgrade has been planned. This is required as FORS2 is using technology and software that is now obsolete and cannot be put and maintained to the standards in use at the Observatory. The project – carried out as a collaboration between ESO and INAF–Astronomical Observatory of Trieste – aims at bringing to the telescope in 2023/2024 a refurbished instrument with a new scientific detector, an upgrade of the instrument control software and electronics, a new calibration unit, as well as additional filters and grisms. The new FORS will also serve as a test bench for the Extremely Large Telescope (ELT) standard technologies (among them the use of programmable logic controllers and of the features of the ELT Control Software). The project aims at minimising the downtime of the instrument by performing the upgrade on the currently decommissioned instrument FORS1 and retrofitting the Mask Exchange Unit and polarisation optics from FORS2 to FORS1.

## 1. THE VERY LARGE TELESCOPE FOCAL REDUCER

### 1.1 FORS project history

FORS[†] (or at least a Focal Reducer) was foreseen from the beginning in the Very Large Telescope (VLT) Instrumentation Plan. The final report[1] of the *Working Group on Imaging and Low-Resolution Spectroscopy* recommended "that one of these instruments should be a set of general-purpose focal reducers at one Nasmyth focus of each of the array elements" (=unit telescopes). At the end, two significant departures from this recommendation were made: a) only two FORS instruments were built and b) they went to the Cassegrain instead of the Nasmyth foci as requested – the Cassegrain focus was not yet foreseen at the time the Working Group issued its recommendation.

The "VLT Instrument Consortium" (VIC) was selected to build two identical copies of the Focal Reducers after a competitive Call for Tender. Its members were Landessternwarte Heidelberg (Project Office, PI I. Appenzeller), Universitäts-sternwarte Göttingen (Co-I W. Fricke) and Universitätssternwarte München (Co-I R. Kudritzki), all in Germany. Bernard Delabre had by then designed for ESO several focal reducers capable of imaging using various filters, and low-resolution spectros-copy using grisms. He made a conceptual design for FORS (FOcal Reducer low dispersion

---

[*] hboffin@eso.org
[†] In the following, we use the names FORS1 or FORS2 when discussing the specific instrument attached to the Very Large Telescope, and FORS for the concept of the instrument or for the upgraded instrument.

Spectrograph) that was accepted and refined by the VIC for their response to the original ESO Call for Tender. The VIC design as finally built accommodated 3 wheels in the collimated beam with altogether 14 broad band filters, 6 grisms and the Wollaston prism, 8 interference filters in two wheels in the converging beam in front of the CCD, and the polarimetry optics to be mounted permanently on each FORS.

The contract was signed in December 1991, Preliminary Design Review was passed in April 1992, Final Design Review was concluded in September 1994, and Preliminary Acceptance Europe of FORS1 took place in July 1998. FORS1 saw First Light on September 15th, 1998, FORS2 on October 29th, 1999. FORS1 entered regular science operations on April 1st, 1999 and was retired on April 1st, 2009 and is now stocked at La Silla. FORS2 entered regular science operations on April 1st, 2000 and has been observing ever since without significant interruption[2,3,4,5].

At *First Light*, both FORSes were equipped with 2K x 2K Tektronix CCDs with 20 μm pixels and identical broadband antireflection (AR) coatings. Quite quickly, in October 2001, a new detector mosaic of two 2K x 4K 15μm pixel MITLL red-sensitive chips was commissioned on FORS2, which is in continuous operation since April 2002, with short interruptions when the "blue" mosaic (by e2v, but having the same format as the MITLL red) is scheduled. The blue mosaic was installed[6] on FORS1 in 2007 and is offered in visitor mode on FORS2 since 2009, when FORS1 was decommissioned. During the time FORS2 was in operation, volume phase holographic gratings (VPHGs) became available that offered higher efficiency and several of them are now in operations[7]. Problems were often encountered with over- and under-exposure when obtaining screen flats with the built-in calibration units and a new External Calibration Unit (ECU) sitting on the telescope's M1 cell instead of on the instrument's top section was designed and built. Similarly, the initial Echelle mode that was present on FORS2 has been decommissioned, and so were also the High-Time Resolution imaging and spectroscopic modes. More information is available at *eso.org/sci/facilities/paranal/instruments/fors.html*.

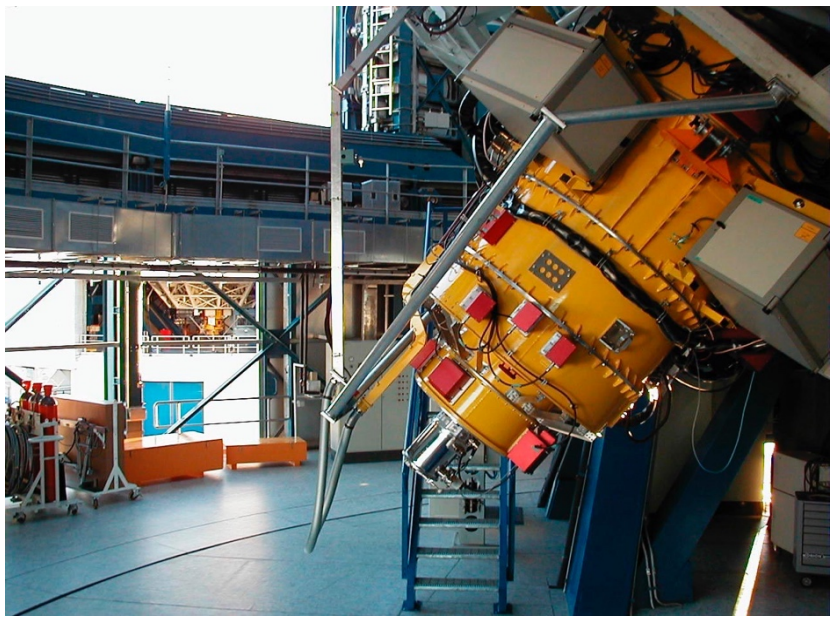

**Figure 1.** This photo shows the twin instruments prior to 2009, FORS2 at KUEYEN (in the foreground) and FORS1 at ANTU, seen in the background through the open ventilation doors in the two telescope enclosures.

## 1.2 FORS2 in the context of the Very Large Telescope

The FORS1 and FORS2 instruments were part of the first generation of VLT instruments. On the VLT, they were the only scientific imagers and low-resolution spectrographs in the visible (until the arrival of VIMOS, which has by now been decommissioned), and the only instruments for absolute polarimetry. Both have motorized multi-object spectroscopy units with 19 slits. FORS2 is unique with its Mask Exchange Unit (MXU) that was not foreseen originally but added to the instrument specification during the Final Design Phase. It was paid for (as well as a field of view increase in both Standard- and High-Resolution modes) by only buying one set of the very expensive polarimetry optics. These can however be used in both FORS1 and FORS2 and indeed were moved to FORS2 in 2009. FORS2 is currently the only visible-light imager (broad- and narrow-band) with a field of view of 7x7 arcmin, the only multi-object spectrograph in the blue and the only polarimeter on an 8-m telescope available in the ESO instrument suite, at least when dealing with faint objects. It is still one of the most demanded instruments in Paranal and most of its capabilities will not be covered by upcoming instruments.

## 1.3 Science goals

When conceived, the main scientific objective of the FORSes was to extend ground-based spectroscopy and photometry to significantly fainter objects than could be reached so far. Among the tasks of FORS1/2 were to be the quantitative analysis of the properties of galaxies at distances up to 10 billion light-years and beyond. Investigations of clusters of galaxies and of single stars in galaxies were to shed light on the nature of the dark matter. Spectroscopy of luminous stars, supernovae and planetary nebulae in remote galaxies would allow us to obtain a better knowledge of the expansion of the universe and would provide information on the origin of chemical elements outside our own Galaxy. One can say that most of such observations have indeed been realised. Over the years, and the availability of new instruments, the science done with FORS2 has, however, shifted.

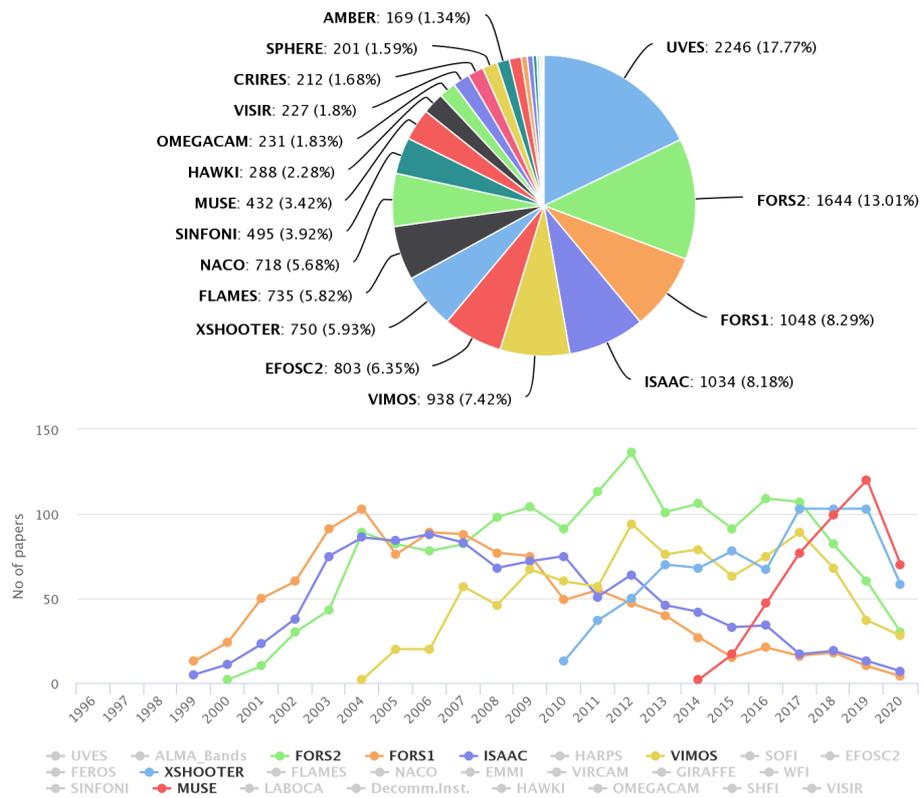

**Figure 2.** (Top) Total number of refereed papers produced by VLT instruments over the years. (Bottom) Number of yearly refereed papers for some of the VLT instruments.

The FORS instruments are among the most successful instruments in Paranal (Fig. 2): FORS1 led to 1048 refereed publications, while for FORS2 this number is 1644 (as of November 2020). Given the competition of other instruments, the number of yearly publications is decreasing. Thus, if in 2017 alone, FORS2 led to 106 refereed publications, putting it in second place of all La Silla Paranal instruments, in 2019, this was reduced to 60, putting FORS2 in fourth place of all VLT instruments, behind UVES, MUSE and XSHOOTER. Quite remarkably, not as long ago as in 2018, there were still 18 refereed papers using FORS1 data, mostly in combination with more recent data from FORS2, although some were using only FORS1 data. An early presentation of the FORS science can be found in [4]. At that time, it was already stated that "*if we look at the number of citations of VLT papers, it is symptomatic that at least one of the two FORS instruments was involved in eight of the ten most cited VLT papers*". Among the most cited FORS2 papers, one finds obviously the spectroscopic study of the GOODS-South field and of the Chandra Deep Field-South, which are based on MXU observations. Looking at the most recent years, the most cited papers are about Ly-alpha emitters in the early universe (MXU), spectro-polarimetry of massive stars, photometric studies of young stellar regions, astrometric studies of brown dwarfs and transmission spectroscopy of exoplanets. This shows that all modes of FORS2 are used and lead to high impact science. Some of the most recent publications coming out from FORS2 still highlight the innovative science done with it: the first interstellar asteroid or comet detected in the solar system, the spectroscopic identification of the gravitational wave source and the first detection of titanium oxide as well as the absolute sodium abundance in the atmosphere of an exoplanet. In 2019, there were two papers published in Science, both about Fast Radio Bursts, showing again how FORS2 is ideal to tackle the most recent hot topics. Over the years, FORS2 (resp., FORS1) led to 29 (14) *Nature* papers and 12 (8) *Science* papers. Apart from the above-mentioned topics, these papers dealt with a variety of topics, including asteroids, binary stars, neutron stars, supernovae, black holes, gamma-ray bursts, and quasars. For a full list of refereed publications based on FORS, we refer to the ESO Telescope Bibliography, at *telbib.eso.org*.

It is interesting to note that much of the current science done with FORS2 differs from what was initially foreseen[8,9] and therefore leads to different requirements[10]. Many current programmes use relatively short exposure times (and some do time monitoring), are mostly photon-noise limited and would benefit from short CCD read-out times. Also, although imaging and spectroscopy is done by default with a 2x2 binning, for spectro-polarimetry, the 1x1 binning is preferred, as it avoids saturation for generally bright sources (polarimetry needing high signal-to-noise ratios) and allows getting a higher spectral resolution. Moreover, the instrument modes being preferably used are changing. Compared to previous years, if the usage of the imaging mode remains stable, both polarimetric imaging and long-slit spectroscopy seem more demanded, while MXU and spectro-polarimetry appear less requested. Here is a list of some of the recent FORS2 science cases and their requirements towards the instrument:

- *Radial velocity measurements of close binary stars*[11]: this requires a signal-to-noise (SNR) of about 50 in 900s exposure at a resolution of R~2500 as the signal shouldn't be smeared out by the orbital motion. This also ideally requires to be able to obtain wavelength calibrations during the night.
- *Exoplanet transmission spectroscopy*[12,13]: stable multi-object spectroscopy of relatively bright stars (photon noise dominated) over the largest field of view (FoV) possible. It is important to have a stable instrument over several hours and to cover the Na I and K I lines, as well as the blue part of the spectrum. Final accuracy reached should be of the order of 100 ppm.
- *Spectro-polarimetry of bright stars*[14,15,16]: the shortest exposure times should be allowed and precise polarimetry should be possible.
- *Detection of faint Lyman alpha from a z=6-7 object*[17,18]: this requires being able to detect a line with an equivalent width of 30 Å and a flux of 3.3 e-20 erg/s/cm2/Å in the wavelength range of 940-1000 nm in a few hours.
- *Extragalactic transients, e.g., the study of gravitational-wave event electromagnetic optical counterparts*[19] *and the study of super-luminous supernovae*[20]: new discoveries in this field are generally of rare (in their volumetric rates) events and are therefore faint, requiring the highest possible instrument sensitivity (a requirement that will only become more pertinent in the Vera Rubin Observatory era). In addition, polarimetric studies of these objects (to probe explosion symmetries) require the highest possible polarisation precision.
- *Solar system transients*[21] *and interstellar interlopers*[22]: the study of comets, asteroids and visits from interstellar objects. Again, instrumental requirements are the highest possible throughput and the highest possible polarisation precision.

## 1.4 Observing Modes

FORS2 is the acronym for FOcal Reducer low dispersion Spectrograph 2, which is a clear understatement. Indeed, the instrument should better be called FORFISMOSP 2, for FOcal Reducer Fast Imager and low dispersion Single and Multi-Object Spectro-Polarimeter 2, given all its observing modes: imaging, spectroscopy (slitless, long-slit and multi-object), and polari-metry, which we describe in more detail here. A schematic view of FORS is shown in Fig. 3.

FORS2 imaging covers a field of view of 6.8x6.8 arcmin (standard collimator, with pixel sizes of 0.126 arcsec) or 4.2x4.2 arcmin (high resolution collimator, with pixels of 0.0632 arcsec), from the *U* to the *Z* band. It is characterized by a superior image quality (in part due to a passive flexure compensation system). The Unit Telescope where it is mounted is furthermore equipped with a Longitudinal Atmospheric Dispersion Corrector (LADC) that is effective up to about 60 degrees away from zenith (i.e., airmass 2). FORS2 can do spectroscopy in three modes, over the wavelength range from 330-1000 nm: classical long-slit spectroscopy (LSS) with slits of 6.8 arcmin length and predefined widths between 0.3 and 2.5 arcsec; multi-object spectroscopy (MOS) with 19 slitlets of 20–22 arcsec length each and arbitrary width (above 0.2 arcsec) created by movable slit blades; and multi-object spectroscopy using masks (MXU) with slitlets of almost arbitrary length, width, shape and angle. There are 15 grisms available with resolutions (for a 1 arcsec slit) from 260 to 2600, which may be combined with three different order-separating filters to avoid second-order contamination. It is also possible to do slitless spectroscopy with FORS2, in MOS mode with all slits open. FORS2 can also perform both imaging and spectroscopic polarimetry (IPOL and PMOS). Both circular and linear polarimetry are possible.

What sets FORS2 currently aside from the other, or planned, VLT instruments is the following:

- Imaging: it is the sole imager in the visible (330-1000 nm), with broadband and narrow band filters, with the possibility to reach a demonstrated 0.18 arcsec resolution in the optical, over a field of view of almost 7x7 arcmin with an 8-m telescope. In the optical, MUSE allows to recreate images, but its FoV is 1x1 arcmin only (in Wide Field Mode) and with a spatial sampling of 0.2 arcsec, while not going as much in the blue as FORS2, and the XSHOOTER acquisition camera also allows to do imaging, in a 1.5x1.5 arcmin FoV but only with the Johnson and SDSS filters, and with a pixel size of 0.18 arcsec. HAWK-I is an imager in the infrared, complementary to FORS2, even though it is now equipped with ground layer adaptive optics to reach a much higher spatial resolution.
- Spectroscopy: long-slit spectroscopy in the visible for faint objects, at low and medium resolution. MUSE provides a slightly higher spectral resolution (R~1800 -3600) and doesn't go so much in the blue (MUSE goes down to 465 nm in the extended mode, up to 930 nm). XSHOOTER has a higher spectral resolution (R~10,000), and covers the full UV-IR range, but cannot observe as faint objects as FORS2 and provides only a short slit (11 arcsec).
- Multi-object spectroscopy: the possibility to use MOS to observe up to 19 objects, or the MXU masks to observe up to ~100 objects across a 6.8x6.8 arcmin FoV, makes FORS2 a unique multiplex low- and medium-resolution spectrograph in the visible at the VLT, especially up to the very blue atmospheric cut-off. KMOS allows up to 24 spectra in the infrared and MUSE allows to do Integral Field Spectroscopy, but over a smaller FoV of 1x1 arcmin, and not in the bluest part. FLAMES/GRAFFE performs medium- and high-resolution spectroscopy with up to 130 fibres in a 30x30 arcmin FoV, but the wavelength range is more limited, and the limiting magnitude is much brighter than with FORS2.
- Polarimetry: the FORS imaging and spectroscopic polarimetry modes are unique at the VLT. While imaging polarimetry is offered by the second-generation instrument SPHERE, the science cases between the two instruments are quite distinct. FORS2 is the only instrument that offers accurate absolute polarization measurements for faint objects at the VLT.

EFOSC2 is the little brother of FORS2 at the NTT in La Silla and can perform imaging (with a 4.1x4.1 arcmin FoV), low-resolution spectroscopy and multi-object spectroscopy, but on a smaller telescope (3.6 m vs. 8.2 m). HARPS at La Silla on the NTT can do high-resolution spectro-polarimetry and is thus only suited for relatively bright targets.

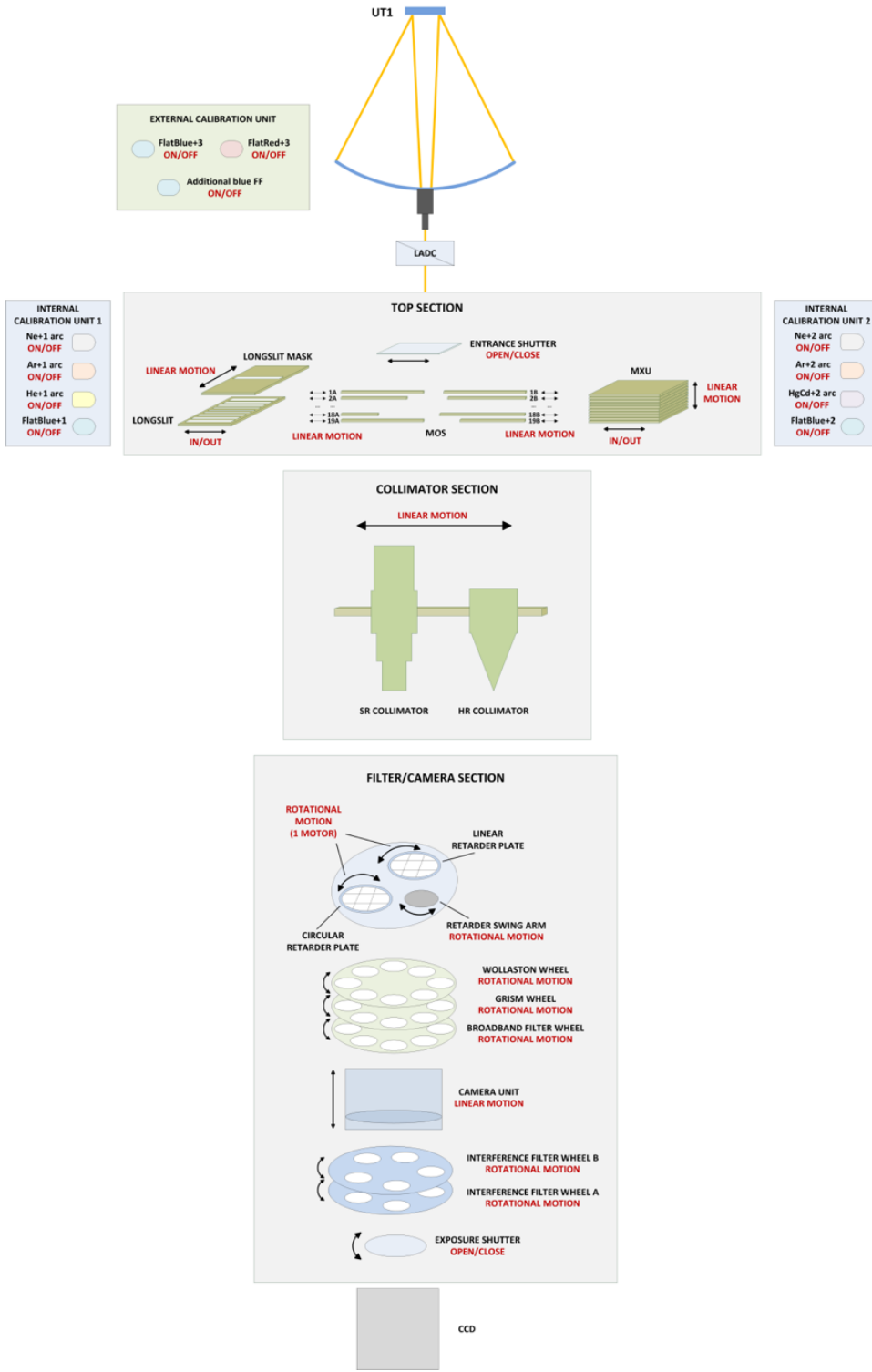

**Figure 3.** Schematic FORS instrument overview. It physically consists of the following parts: External calibration unit; Top section (entrance shutter, a Multi-Object Spectroscopy unit with 19 movable slits, a long slit mask unit with 9 slits and a mask exchange unit for MOS spectroscopy holding up to 10 masks); Internal calibration units, equipped with arc lamps; Collimator section, which contains the Standard Resolution and High Resolution collimators and the electronic cabinets; Filter/camera section, harbouring the retarder plate mosaics swing arm with the linear and circular retarder plate mosaics, the camera focusing stage, the exposure shutter, five filter wheels containing the Wollaston prism, the grisms and the broadband and interference filters; and finally the scientific detector.

## 1.5 Current control architecture

*Control Software*

The current FORS2 control architecture is based on the VLT Software[23] (VLTSW), a distributed system connecting a set of workstations, dedicated to high level operations, and Local Control Units (LCUs), dedicated to the control of sub-systems hardware. The chosen programming language for the workstation applications (running Linux operating system) is mainly C++, except for the Graphical User Interfaces, as well as High-Level Operations Software (e.g., BOB, the Broker for Observation Blocks and the instrument templates), which are based on Tcl/Tk. The VLTSW are released by ESO on periodical basis (the official release at the time of writing is VLTSW 2019). In the latest releases, the support of programmable logic controllers (PLC) for the hardware control has replaced and substituted the old VME VxWorks-based LCUs.

Several algorithms implemented in the control software (especially those dedicated to the optical alignments) are based on ESO-MIDAS[24], a system which provides general tools for image processing and data reduction, but which is not maintained by ESO anymore.

Figure 4 shows the current FORS2 control architecture. The control software, running on the instrument workstation, communicates with the software that controls the telescope and the scientific detector and with the three LCUs controlling the instrument hardware. During the observing run, the Broker for Observation Blocks (BOB) executes the templates contained in the Observation Blocks, the fundamental scheduling units of VLT science operations. Each template consists in general of a sequence of commands to be sent to the Observation Software (OS), the software subsystem in charge of managing the whole scientific exposure. Then OS translates the incoming commands into further commands to be sent to the different subsystems (telescope, scientific detector and instrument hardware).

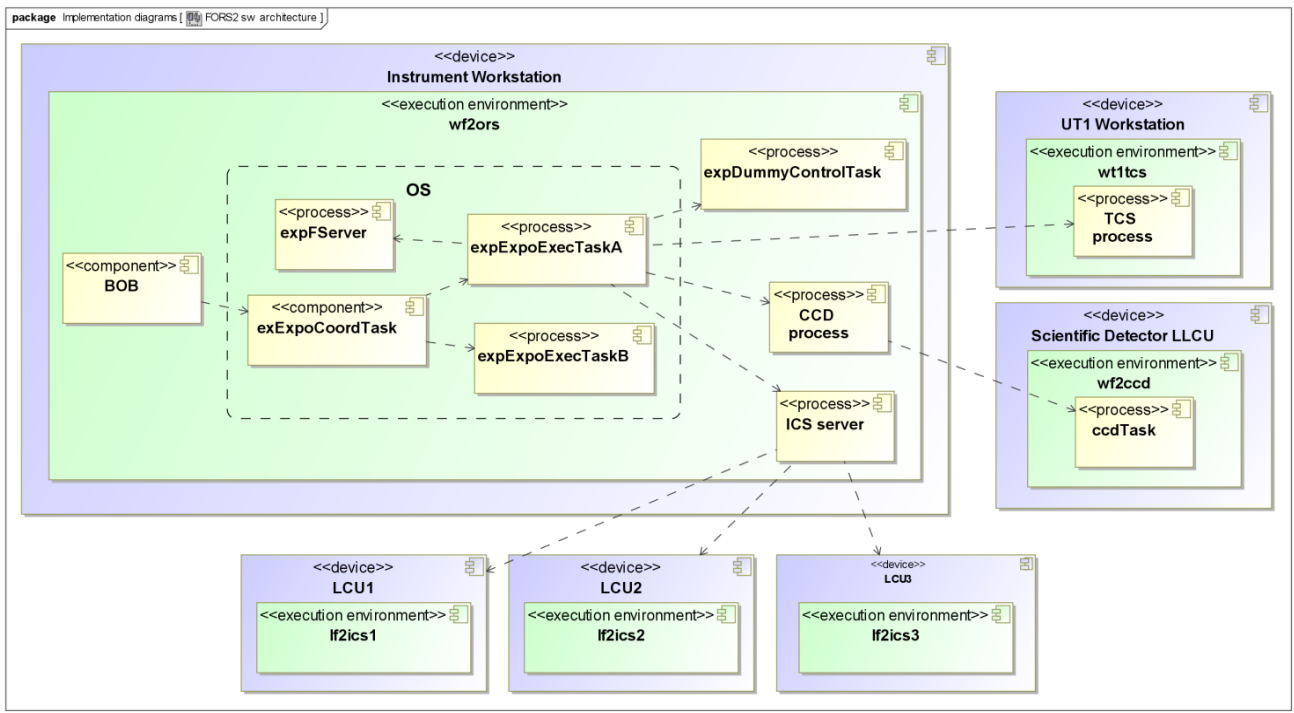

**Figure 4.** Current FORS2 control architecture (see text for details).

*Control electronics*

The control electronics of FORS2 is based on the VME-based Local Control Units (LCUs), since this was the standard for the control electronics of ESO instruments at the time of the FORS2 construction, following a similar trend in other big scientific experiments. The crates are equipped with specialized I/O and motion control boards, some of them being custom made. This architecture has now started to be difficult to support and maintain and a replacement is needed. PLC and commercial-off-the-shelf based control electronics building blocks have superseded the old VME based systems in the new designs. The FORS2 control hardware is split into three parts, each one mounted in a 19" system cabinet, plus a fourth cabinet for the scientific detector system.

## 2. THE FORS-UP PROJECT

Due to its versatility, the ESO Scientific Technical Committee has identified FORS2 as a highly demanded workhorse among the VLT instruments that shall remain operative for the next 15 years[25]. However, as shown above, the current science done with FORS2 differs from what was initially foreseen. Many current observing programmes use relatively short exposure times, are photon-noise limited and therefore, beside efficiency, the read-out noise of the CCD become an issue. An upgrade with a 4k x 4k broadband detector shall improve the operations of the instrument, eliminating also the need for the exchange of the red or blue detector systems on the instrument and removing the current gap between the two chips. Moreover, both software and electronics controlling the instrument have been developed at the end of the 90's and several control parts are obsolete and not supported anymore by vendors. *All these considerations led to the FORS-Up project*. The main goals of the project, besides the upgrade of the FORS2 scientific detector, is the upgrade of the instrument control software and electronics to the standard developed for the forthcoming Extremely Large Telescope (ELT), and some additional optical components. In addition, the upgrade aims at changing the full control electronics, motors, sensors and all of the cablings.

### 2.1 New detector

A much-required upgrade is that of the science detector. The upgrade foresees to replace the two, red and blue, detectors that are each composed of two 2kx4k chips, by a single 4kx4k chip that should allow an almost as good coverage of both the blue and red parts of the spectrum. The proposed chip is a variant of the CCD used in the MUSE project, the e2v CCD231-84. The differences with respect to the MUSE devices are:

- The MUSE chip has a graded anti-reflection (AR) coating optimised for the fixed spectral format that the instrument's spectrographs record. This graded coating is not adequate for FORS, given the different modes of operations in the instrument, imaging with different scales and spectrograph, recording spectra with different resolutions and with different wavelength bands. A homogeneous AR coating is required for the FORS Upgrade.
- Fringing is a critical aspect that shall be considered for the new device to be used, as FORS should observe also in the very red. e2v offers a fringe suppression process in the fabrication, which greatly reduces (if not eliminates) this effect. Fringe suppression process will be required in the device for the FORS Upgrade.

The current FORS2 cryostat and optical mount are compatible with the new selected detector. This simplifies the design, the development and the operation of the FORS2 upgrade.

ESO has developed the New General Controller (NGC[‡]), which is the new standard controller for the second generation of VLT instruments. The NGC offers improved performance compared to the FIERA controller currently used. It has smaller size, lower weight and less heat dissipation. It includes the shutter control, with the possibility to interface with TIM boards for readout synchronization (not supported in FIERA). Moreover, the NGC base software allows the implementation of more complex readout patterns. In the framework of the new ELT technologies, ESO is currently developing a new version of its NGC (NGC II) and it is expected that this is the version that will be used with the new detector on the upgraded FORS.

---

[‡] https://www.eso.org/sci/facilities/develop/detectors/controllers/ngc.html

## 2.2 Upgrade plan

The best option for the upgrade is to work on the decommissioned FORS1 instrument, by recovering it in Europe. The full availability of the instrument structure and mechanics should allow to work on full refurbishment and exchange of every control system component, without impacting operations of FORS2. FORS1 lacks the MXU part and the polarisation optics, but still has all the other parts that could be used for full system testing and validation before reintegration at VLT. Only the top section of FORS2 and the polarisation optics shall at that time be recovered, upgraded and moved to the refurbished FORS-Up instrument.

One major step forward in this proposed baseline is the use of brushless DC-motors, chosen from the Beckhoff catalogue. This choice follows the modern trends in control systems, that should grant a long-term support for the components employed.

The baseline Beckhoff configuration has been tested on FORS1 at La Silla in February 2020. This first fit check confirms that the Beckhoff configuration can be mechanically integrated on FORS and that the selected motors are fulfilling the function. This test was performed without control and will be repeated in full details when FORS1 will be in Europe.

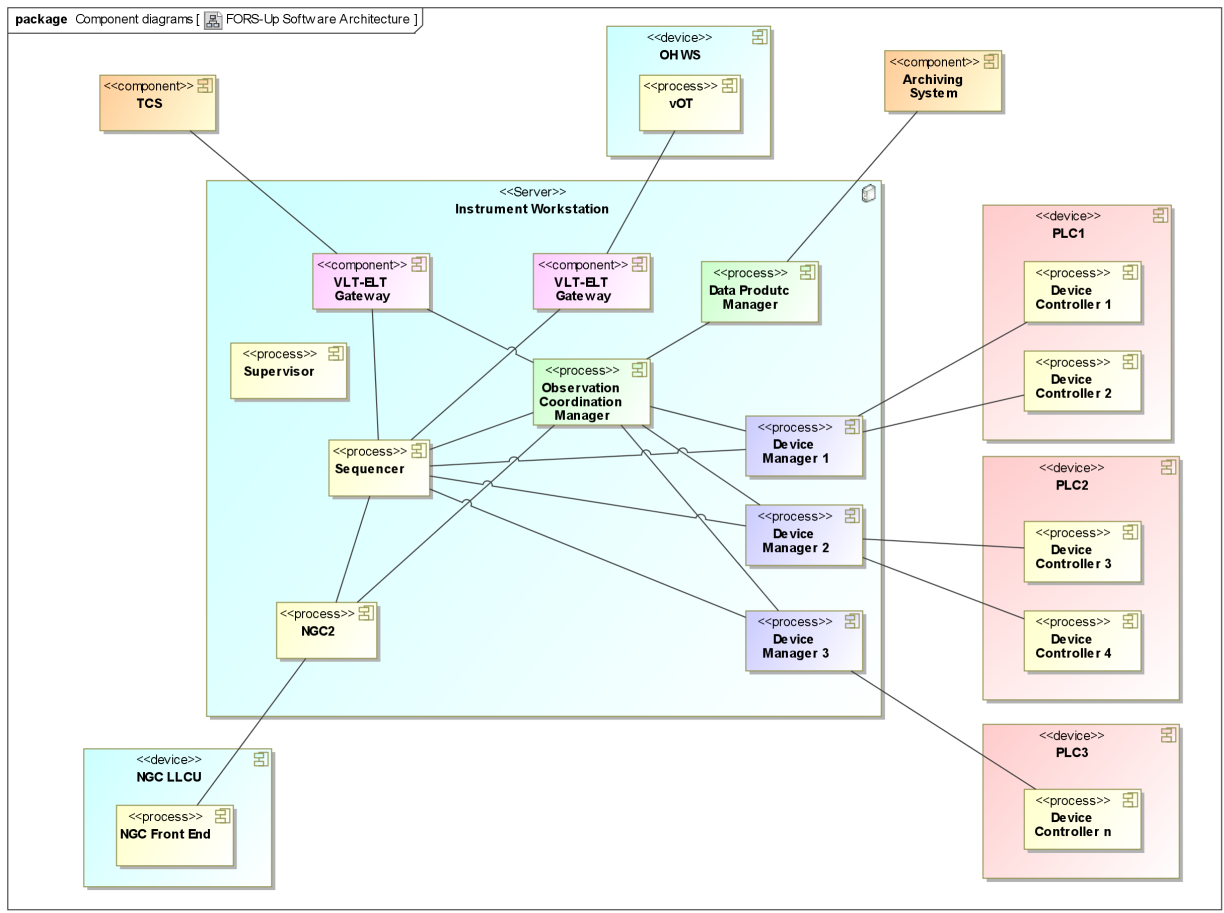

**Figure 5.** Software architecture based on the ELT ICS Framework foreseen for the upgraded FORS.

## 2.3 Control architecture

Although it was initially foreseen that the FORS-Up control software will be based on the latest VLTSW release, following the recent developments in software and hardware technologies for the ELT, it has been decided that the upgrade of the FORS control software will adhere to the ELT standards and will be based on the ELT Instrument Control Software (ICS) Framework[26]. *It will be the first ESO* *instrument to do so!* The aim is to develop a full-fledged control system able to efficiently fight hardware obsolescence, offer modern software tools, lower costs, integration and maintenance efforts and easy installation.

From the hardware point of view, the most notable modification is the adoption of Beckhoff PLCs, which replace the VMEs. The original design topology has been kept, where in each of the three cabinets a PLC control unit finds its place. The I/O modules are either connected on the CPU itself or decoupled using EtherCAT bus and couplers.

The proposed baseline design does not foresee changes in the mechanical architecture, but rather a change of the motor technology by employing Beckhoff brushless motors that are conceived to work in conjunction with the control modules of the same manufacturer and integral part of the PLC system.

The FORS-Up software architecture based on the ELT ICS Framework is shown in Figure 5. In this architecture the Sequencer executes the Observation Blocks and interfaces directly with the instrument subsystems (Observation Coordination Manager which coordinates the data acquisition, NGC II for the control of the scientific detector and Device Managers for the control of the instrument hardware). Each Device Manager, at start-up, creates the instances of the Device Classes responsible for the communication with the instrument devices. The communication with the Beckhoff PLCs is performed through the ELT standard OPC-UA communication protocol. In the FORS-Up control software design, almost all the devices will be controlled by standard Device Classes provided by ESO in the framework of the ELT ICS architecture. The only non-standard devices, which will require a dedicated design and implementation, are the MOS slits (due to the fact that their movement must be synchronized) and the exposure shutter (in operations, it has to be controlled by the NGC II, see above, running on a dedicated machine, and in maintenance by the Device Manager running on the instrument workstation). A dedicated ELT-VLT gateway software, developed by ESO, will allow the communication between the instrument control software based on the ELT standards and the VLT software subsystems, e.g., the Telescope Control Software (TCS)

Finally, the ESO-MIDAS routines used by the OS alignment algorithms will be re-implemented using the Online Data Processing (ODP), a component of the ELT ICS Framework which aims to provide a flexible data processing toolkit.

### 2.4 New grisms

In addition to the main aims of the upgrade, which are the replacement of the detectors, of the motors, sensors and cabling, and the upgrade of the control software, the Phase A of the FORS-Up project also identified the usefulness to add new grisms and new filters to the instrument. This need will be established more precisely during the Phase C of the project.

Concerning grisms, three elements are identified for a possible upgrade, one with low dispersion to replace the current GRIS 600B+22, and two new ones with moderate dispersion (covering the Na I Lines and the K I line, resp.).

The main specifications are reported in Table 1.

Table 1. FORS new grism typical performances.

| Property | GRISM 600B | Na line @ 580 nm | K line @ 770 nm |
|---|---|---|---|
| Central wavelength ($\lambda_c$) (nm) | 460 | 580 | 770 |
| Bandwidth (nm) | 330 – 620 | 524 - 640 | 695 – 849 |
| Line density (l/mm) | 659 | 1464 | 1181 |
| Incidence angle in air (°) | 8.7 | 25.12 | 27.04 |
| AR coating (% avg) | <1% | <1% | <1% |
| Order sorting filter | N/A | N/A | Es. SHOTT RG 630 |
| Resolution @$\lambda_c$ 1" slit | 780 | 2500 | 2500 |
| External Prisms Material | Fused Silica | PBH71 | LASF35 |

The diffraction efficiency of the VPHGs have been predicted by means of RCWA (Rigorous Coupled Wave Analysis) simulations. Those simulations have been performed by our colleagues at INAF Brera and give an estimation of the throughput of the VPHGs without additional losses (such as absorption, reflections and scattering).

### 2.5 New filters

Currently, FORS2 can accommodate:

- Broadband filters in the collimated beam: There are three wheels in the collimated beams with seven positions available per wheel. A number of positions are usually occupied by grisms and the Wollaston prism. Broadband filters have a free diameter of 138 mm and are bayonet mounted.
- Interference filters in the converging beam of the camera: There are two wheels in the converging beam with together eigth available positions. Interference filters have a size of 115 mm and are bayonet mounted.

During Phase A, it became apparent that the lack of SDSS (or equivalent) filters for FORS2 is rather surprising. With the recent advent of the large surveys, such as SDSS and Pan-STARRS, these filters have become widely adopted by many facilities (including the more recent ESO instruments, such as GROND, VST/OmegaCam, the X-shooter acquisition camera). The Vera Rubin telescope will use 6 filters, which do not exactly match the SDSS bands[§], and more generally, in the field of transient astronomy, it is becoming more and more common to create light curves using such filters. For this reason, the lack of these filters at both NTT (for ePESSTO) and at FORS2 may seem unsatisfactory. Transformations between the two systems are of course possible but require extra work and always leave systematic or unaccounted sources of error in the process. Absolute calibration has also become easier for the SDSS filter set thanks to the now fairly large coverage from Pan-STARRS over 3/4 of the sky and in the future with that offered by the Vera Rubin Observatory (though, it is probably not as good as the one that can be achieved with Stetson standard stars, when done properly, in photometric nights). Thus, the Phase A has identified the need to consider whether a set of SDSS or Vera Rubin Observatory filters for FORS should be provided. The decision to proceed with the procurement of additional set(s) of filters for FORS2 will be taken in the next Phase C of the project.

### 2.6 Fighting stray light

The flat fields of the low resolution grisms covering the range below 3900 Å show significant stray light from the red flat field lamp (Fig. 6), which corrupts the flat field correction and thus also the flux calibration at those wavelengths. This stray light can be seen in both FORS1 and FORS2 data back to at least 2002.

The origin of this stray light has been identified, using Zemax simulations, as due to the zero-order light reflected by the grisms. Given the FORS optical configuration, the optical beam that enters the camera is collimated. Moreover, the FORS field size is 6.83 x 6.83 arcmin (i.e. Ø 9.66 arcmin) on sky and the pupil diameter (for the SR collimator) is Ø 90 mm. Thus, the design acceptance angle for the camera is 14.52º (i.e., ±7.26º). Using raytracing, with a parallel beam with incidences above 7.26º, it was then noticed that when incidence is at 8.5±0.5º the beam illuminates the edge of the first camera lens and could produce an illumination pattern on the detector. The angle corresponds to the diffraction angle for zero order for the GRIS_600B white light and as such cannot be filtered out by any specific optical filter.

It is possible to mask the edge of the first camera lens by preparing a baffle with a clear aperture of Ø 132mm that avoids the illumination of the edge of the lens. This will be managed as part of the FORS-Up activities.

---

[§] https://speclite.readthedocs.io/en/latest/filters.html

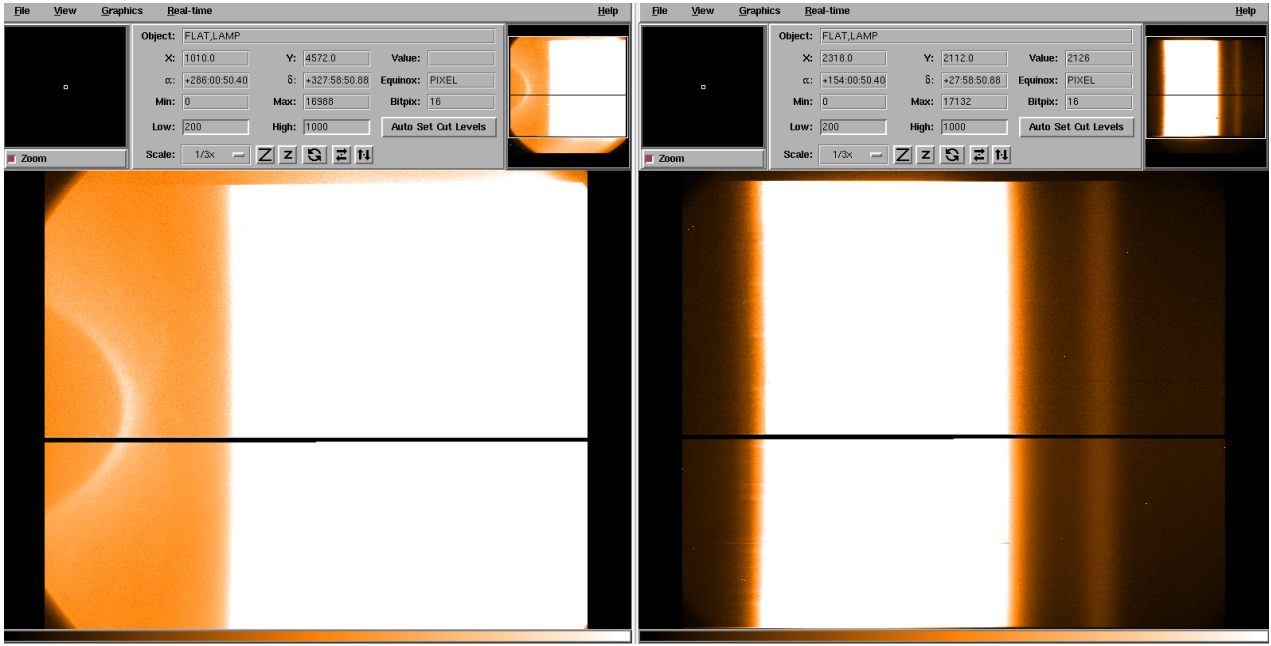

**Figure 6.** Raw flat field data (both detectors combined) observed with the LSS_1.0 slit for GRIS_600B using only the red flat field lamp (left) and only the blue flat field lamp (right). Both images are shown with a dynamic range of 200 ADU (bias level) to 1000 ADU. Clearly, the blue flat field lamp provides negligible flux at the blue end, while the red lamp creates stray light in that region.

## 3. CONCLUSION

The Phase A of the FORS-Up project has been completed in 2020 and the project given green light to proceed. It has identified the various items that will require an upgrade, as well as some further items to be considered. It also identified the best scenario for proceeding with the project, that should ensure the highest on sky availability of FORS2: full retrofit of FORS1 in Garching and commissioning at the telescope, while decommissioning FORS2.

This encompasses the following stages:

- Design in Europe
- FORS1 back in Europe
- Full cleaning, inspection and full refurbishment of FORS1 for performance
- Functional test of ELT standard control electronics
- Deployment of ELT S/W framework on FORS1
- Acceptance test of SW upgrade on ELT Control Model with nominal ELT standards hardware (Beckhoff)
- Acceptance performance test at system level on FORS1 (Performance)
- FORS1 back to Chile
- First commissioning of FORS1 with MOS only; FORS1 starts operations.
- Decommissioning of FORS2
- Refurbishment of MOS/MXU top section (6-month downtime of MXU and polarisation)
- Retrofitting of polarisation optics from FORS2 to FORS1
- Exchange of FORS1-MOS section with refurbished MOS-MXU
- Second commissioning on UT

This should extend the lifetime of FORS by 15 years and it is expected that the refurbished FORS will be on sky in 2023 or 2024.